# Estimation of Field Inhomogeneity Map Following Magnitude-Based Ambiguity-Resolved Water-Fat Separation

Alexandre Triay Bagur[1,2*], Darryl McClymont[2], Chloe Hutton[2], Andrea Borghetto[2], Michael L Gyngell[2], Paul Aljabar[2], Matthew D Robson[2], Michael Brady[2], Daniel P Bulte[1]

[1]Department of Engineering Science, University of Oxford, Oxford, UK

[2]Perspectum Ltd, Oxford, UK

* Corresponding author. Email: alexandre.triaybagur@eng.ox.ac.uk

**Running Title:** Field map Estimation Following Magnitude Water-Fat Separation

**Word Count:** 4480

**Data Availability Statement:** Publicly available 1.5 T data from the UK Biobank study was analyzed in this work. The UK Biobank datasets are available to researchers through an open application via https://www.ukbiobank.ac.uk/register-apply/. The additional 3 T dataset used in this study was de-identified, and code and data were uploaded to Zenodo under https://doi.org/10.5281/zenodo.6652300.

**Conflict of Interest Statement:** ATB, DM, CH, AB, MLG, PA, MDR, MB are employees of Perspectum and hold shares and/or share options. DPB declares no potential conflict of interest.

**Acknowledgement:** The authors thank the Engineering and Physical Sciences Research Council (EPSRC) for funding via a doctoral studentship (project reference 2280970). This research has been conducted using the UK Biobank Resource under application 9914. This publication is our view, and the Executive Agency for Small and Medium-sized Enterprises is not responsible for any use of the information herein.

# Abstract

**PURPOSE**: To extend magnitude-based PDFF (Proton Density Fat Fraction) and $R_2^*$ mapping with resolved water-fat ambiguity to calculate field inhomogeneity (field map) using the phase images.

**THEORY**: The estimation is formulated in matrix form, resolving the field map in a least-squares sense. PDFF and $R_2^*$ from magnitude fitting may be updated using the estimated field maps.

**METHODS**: The limits of quantification of our voxel-independent implementation were assessed. Bland-Altman was used to compare PDFF and field maps from our method against a reference complex-based method on 152 UK Biobank subjects (1.5 T Siemens). A separate acquisition (3 T Siemens) presenting field inhomogeneities was also used.

**RESULTS**: The proposed field mapping was accurate beyond double the complex-based limit range. High agreement was obtained between the proposed method and the reference in UK. Robust field mapping was observed at 3 T, for inhomogeneities over 400 Hz including rapid variation across edges.

**CONCLUSION**: Field mapping following unambiguous magnitude-based water-fat separation was demonstrated in-vivo and showed potential at high field.

# Keywords

# Introduction

Chemical-shift encoded (CSE) MRI water-fat separation methods have emerged as non-invasive tools for proton density fat fraction (MRI-PDFF or PDFF) and $R_2^*$ ($1/T_2^*$) quantification. PDFF has been applied successfully in non-alcoholic steatohepatitis (NASH) drug trials as a replacement for a liver biopsy for the detection of steatosis[1]. Beyond the liver, PDFF has been proposed as an imaging biomarker in the heart, muscle, pancreas, kidney, and adipose tissue deposits[2,3]. $R_2^*$ has shown utility in the quantification of liver iron content (LIC)[4].

The majority of advanced CSE algorithms, such as Iterative Decomposition of Water and Fat with Echo Asymmetry and Least-Squares Estimation (IDEAL)[5] and Variable Projection (VARPRO)[6], are complex-based, in the sense that they require both the MRI magnitude and phase images as input. Complex-based CSE estimates fat and water proportions indirectly through the estimation of $B_0$ field inhomogeneity, conventionally referred to as the "field map". Each field map value leads to a unique solution for water and fat proportions within a voxel[5–7]. However, this implies dependence of water and fat estimation on accurate field map estimation, which is a nonconvex optimization problem with many local minima[7,8]. Erroneous field map convergence may lead to incorrect PDFF and $R_2^*$ quantification, and may manifest as fat-water swaps in the PDFF images[7]. Spatial smoothness constraints are often imposed on the field map to regularize the optimization, for instance through using region-growing[7,9,10], multiresolution techniques[9,11], Markov Random Fields (MRF)[6,12], or graph cuts[8,13–16]. These methods have been shown to work well under a range of scanning conditions and anatomies[14,16]. However, the imposed smoothness constraints may be inappropriate, for instance depending on the magnetic susceptibility of the object, or in transition regions with high field inhomogeneity, causing over-smoothing of PDFF and compromising any downstream quantification.

CSE methods that use only the magnitude of the MRI source images, which we refer to as magnitude-based, have also been proposed[17]. Magnitude-based methods do not require field map estimation prior to water-fat separation and are unaffected by errors in phase images due, for example, to eddy currents. In practice, complex-based methods are often subject to a final magnitude fitting step to correct phase errors in water and fat images, an approach that has been dubbed "hybrid" estimation[18]. Magnitude-based methods optimize water and fat (and thus PDFF) directly but have tended to suffer from water-fat ambiguity, meaning they could not identify the dominant species within a voxel (water or fat), which limited the dynamic range of PDFF to 0 to 50%[1,17,19]. A recently proposed method, MAGnitude-Only (MAGO)[20], exploits the spectral complexity of fat to unambiguously identify the dominant species and thus estimate PDFF over the entire 0-100% range[20]. MAGO has shown to give excellent accuracy and reproducibility across manufacturers and clinical field strengths, in simulations, phantoms and in-vivo studies[20].

Though, under the MAGO framework, field mapping is not required for full-range PDFF and $R_2^*$ mapping, the field map may still be useful to assess image quality or for other downstream applications such as distortion correction and quantitative susceptibility mapping (QSM)[21,22]. QSM methods, for example, decompose the field map into a background component and a local component, and have been successfully applied in the abdomen for the measurement of liver iron[21].

In this study, we demonstrate field mapping using complex-valued data following magnitude-based water-fat separation and $R_2^*$ estimation, where the water-fat ambiguity has been resolved previously using MAGO. The method is described using theory and simulations. We show that the calculated field map may be used to refine PDFF and $R_2^*$ under noise. The method is first validated against a state-of-the-art complex-based method in a cohort from the UK Biobank imaging sub-study (1.5 T), which presented low field inhomogeneities. The method is then applied to one 3 T acquisition with high field inhomogeneities, including rapid field transitions.

## Theory

### Signal Models

PDFF fitting conventionally uses the following generalized signal model at each voxel, which is not made explicit to simplify the notation:

$$s_n = (\rho_W + \rho_F C_n)\, e^{i(2\pi f_B t_n + \phi_0)} e^{-R_2^* t_n} \qquad \text{Equation 1}$$

where $s_n = s(t_n)$ is the observed signal at echo time $t_n$, where there are a total of $N$ echoes $[t_1, t_2, \ldots, t_N]$, $\rho_W$ is the unknown water magnitude and $\rho_F$ is the unknown fat magnitude, and $f_B$ is the frequency shift due to local field inhomogeneities or "field map" ($2\pi f_B$ is sometimes expressed as $\psi$). The term $C_n = \sum_{p=1}^{P} \alpha_p e^{i2\pi f_p t_n}$ addresses the multi-frequency nature of the fat species, where $\alpha_p$ and $f_p$ are the relative amplitudes and relative frequencies of the spectral model of fat with $P$ peaks. The 6-peak liver model was used in this work[23]. This signal model is phase-constrained[24,25], meaning that water and fat share the same initial phase $\phi_W = \phi_F = \phi_0$, which is a reasonable assumption in spoiled gradient echo acquisitions[26], and provides noise performance advantages[24].

When fitting only the magnitude of the signal, the model becomes

$$|s_n| = |\rho_W + \rho_F C_n|\, e^{-R_2^* t_n} \qquad \text{Equation 2}$$

where $|x|$ denotes the modulus. Note that the exponential term contributes only to the phase information, $e^{i(2\pi f_B t_n + \phi_0)}$, has unit modulus, and is dropped. It follows that magnitude-based methods avoid computing both the field map $f_B$ and the initial phase term $\phi_0$. Finally, PDFF may be computed using the water and fat magnitudes using the ratio

$$\text{PDFF} = \frac{\rho_F}{\rho_W + \rho_F} \times 100 \text{ (\%)} \qquad \text{Equation 3}$$

## Magnitude Fitting with Resolved Water-Fat Ambiguity

In MAGO, at each voxel the magnitude signal equation above is solved twice using nonlinear optimization from two different starting points, one near 0% PDFF, the other near 100% PDFF[20], obtaining the candidate solution sets $\{\rho_W, \rho_F, R_2^*\}_1$ and $\{\rho_W, \rho_F, R_2^*\}_2$. The solution 'chosen' by MAGO at each voxel, $\{\rho_W, \rho_F, R_2^*\}$, is the candidate solution set with lower associated residual sum of squares (better fit). The 'alternative' solution is the one with higher associated fitting residual (worse fit), where fat and water will mostly 'swapped'. Chosen and alternative sets are shown for one subject in Figure 1. The method enables accurate magnitude-based PDFF and $R_2^*$ estimation.

The 'alternative' solution may be kept to further refine PDFF and $R_2^*$, as described in the next sections.

## Proposed Voxel-Independent B$_0$ Field Estimation using Full-Range PDFF and $R_2^*$

The chosen MAGO solution, $\{\rho_W, \rho_F, R_2^*\}$, may be used to guide field map $f_B$ (and initial phase $\phi_0$) estimation using the source complex-valued data. To this end, we explore whether prior knowledge about water, fat and $R_2^*$ may be useful during field map estimation.

For given values of water, fat and $R_2^*$, the following terms of the complex-based signal model in Equation 1 are known: $\varrho_n \equiv (\rho_W + \rho_F C_n)$ and $R_n \equiv e^{-R_2^* t_n}$, which may be rearranged to yield the simplified expression:

$$\frac{s_n}{\varrho_n R_n} = e^{j(2\pi f_B t_n + \phi_0)}$$

This implies that the field map $f_B$ and the phase offset $\phi_0$ can be estimated given $\rho_W$, $\rho_F$, $R_2^*$ and using the input complex-valued data $s_n$. Taking the phase from both sides,

$$P_n = arg\left(\frac{s_n}{\varrho_n R_n}\right) = 2\pi f_B t_n + \phi_0 + 2\pi k \qquad\qquad \text{Equation 4}$$

with $k$ an integer, then the estimation may be formulated as a linear equation in matrix form,

$$\mathbf{P} = \begin{bmatrix} P_1 \\ \vdots \\ P_N \end{bmatrix} = \begin{bmatrix} 2\pi t_1 & 1 \\ \vdots & \vdots \\ 2\pi t_N & 1 \end{bmatrix} \begin{bmatrix} f_B \\ \phi_0 \end{bmatrix} = \mathbf{A}\,\boldsymbol{\varphi} \qquad\qquad \text{Equation 5}$$

where $\boldsymbol{\varphi} = [f_B, \ \phi_0]^T$ are unknowns, and $\mathbf{A}$ is a $N \times 2$ matrix where the first column vector contains the echo times and the $2\pi$ term, $[2\pi t_1, 2\pi t_2, \ldots, 2\pi t_N]^T$, and the second column contains ones $[1, 1, \ldots, 1]^T$. This makes for a computationally efficient estimation.

For each voxel in the source image, the proposed algorithm first creates the $\varrho_n$ and $R_n$ terms using the MAGO chosen solution set $\{\rho_W, \rho_F, R_2^*\}$, the echo times and the 6-peak spectral model of fat. Then, the input complex-valued data $s_n$ is divided by the $\varrho_n R_n$ term, $s_n/(\varrho_n R_n)$. The phase of the output is taken and unwrapped in *the echo times' dimension* (rather than in the spatial dimensions), by changing absolute jumps greater than $\pi$ to their $2\pi$ complement starting from the first echo time. The echo times' matrix $\mathbf{A}$ is then defined and used to find a least squares solution to the linear equation system in Equation 5.

For correction in the presence of bipolar gradients that create an additional phase shift to the signal $(-1)^n \theta$[27], one may reformulate Equation 5 to estimate three unknowns, $\boldsymbol{\varphi}' = [f_B, \ \phi_0 - \theta, \ \phi_0 + \theta]^T$, namely the combination of the common phase offset of water and fat, $\phi_0$, and a phase offset due to bipolar gradient effects, $\theta$, with opposite sign for odd and even echoes. In this scenario, the echo times matrix $\mathbf{A}'$ may be defined as a $N \times 3$ matrix as follows (for even $N$):

$$\mathbf{A} = \begin{pmatrix} 2\pi t_1 & 1 & 0 \\ 2\pi t_2 & 0 & 1 \\ \vdots & \vdots & \vdots \\ 2\pi t_N & 0 & 1 \end{pmatrix}$$

This approach enables parameter sharing of the bipolar gradients term and the initial phase offset, which may be disentangled further into $\phi_0$ and $\theta$. Disentangling into $\phi_0$ and $\theta$ was out of the scope of this work as these two terms are often considered unimportant[28].

We may additionally estimate $f_B$ and $\phi_0$ using this approach for the MAGO *alternative* solution set as well, which may provide a basis to refine the MAGO PDFF and $R_2^*$ maps. One proposed implementation of such refinement is described in the Methods section.

## $B_0$ Field Cost Function: $R(f_B)$

The overall residual of the signal model may be plotted against the field map, for visual assessment of the cost (or loss) function $R(f_B)$[7–9]. For voxel-independent iterative complex-based estimation, for instance the IDEAL approach[5] with $R_2^*$ decay and multi-peak fat spectrum modelling[29], the field map $f_B$ may be estimated by minimizing the residual loss function $R_C(f_B)$:

$$R_C(f_B) = \|[R_1, R_2, \dots, R_N]\|_2^2 = \sum_{n=1}^{N}(s_n - \widehat{s_n})^2$$

where $\widehat{s_n}$ is the noiseless signal equation using the estimated values $\{\widehat{\rho_W}, \widehat{\rho_F}, \widehat{f_B}, \widehat{R_2^*}\}$. For our proposed estimation method, where full-range magnitude-based estimates are used as inputs, the residual loss function to minimize is based on Equation 5:

$$R_M(f_B) = \|\mathbf{P} - \mathbf{A}\,\boldsymbol{\varphi}\|_2^2$$

Figure 2 plots the cost functions for the complex-based formulation, $R_C(f_B)$, and for our proposed formulation, $R_M(f_B)$, within a voxel; the former replicates Figure 1a in Yu et al.[7], but with the addition of $R_2^*$ decay, 6-peak fat spectrum modelling[23], and evenly spaced echo times (3 echoes, TE$_1$=2 ms, $\Delta$TE=2 ms). For the particular case of uniformly spaced echo times, the cost function is periodic with period $1/\Delta$TE[8,9]. The loss function curves were simulated over the field map range -1000 to 1000 Hz at 1.5 T (imaging frequency = 64 MHz), using the true values of water, fat and $R_2^*$. The cost functions were plotted for varying water:fat proportions, namely 4:1, 2:1, 1:1, and 1:2.

Voxel-independent IDEAL estimation wrongly converges to a local minimum ('aliased' solution) when the true field map value is beyond approximately $\pm\Delta f/2$, where $\Delta f$ is the off-resonance frequency of the main fat peak relative to water ($\Delta f \approx 220$ Hz at 1.5 T), as described previously[7,8]. The voxel-independent VARPRO implementation has been reported to produce equivalent results to voxel-independent IDEAL, because ambiguities are dealt with by forcing the field map to be within $\pm\Delta f/2$[8]. Such erroneous convergence may lead to fat-water swapping in the PDFF map[7].

Conversely, the proposed field mapping converges accurately with starting estimates beyond the $\pm\Delta f/2$ range for the same simulated data. Note that our formulation reduces the number of local minima in the search space compared to complex-based field mapping methodologies such as IDEAL, which need to deal with ambiguities relating to water and fat proportions. The proposed formulation is independent of water:fat proportion because the water-fat ambiguity is dealt with at the MAGO stage, and the water and fat signals are demodulated from the source data before field mapping.

# Methods

## MRI Data

Datasets from N=152 nominally healthy UK Biobank (www.ukbiobank.ac.uk) volunteers were gathered, obtained with a Siemens Magnetom Aera 1.5 T scanner (Siemens Healthineers, Erlangen, Germany) using a single-slice, 10 mm slice thickness, 6-echo (TE$_1$=1.2 ms, $\Delta$TE=2 ms) gradient-recalled echo (GRE) protocol designed to minimize T1 bias (5° flip angle). UK Biobank is approved by the North West Multi-Centre Research Ethics Committee, and prior written consent was obtained from all participants. Liver segmentation masks were available from these subjects, that had been obtained using a liver segmentation model[30].

One other dataset was gathered with a Siemens Prisma 3 T scanner from a healthy volunteer to test the method under more challenging field inhomogeneities. Informed consent was obtained from the participant. The scan comprised thirty-two slices including the dome of the liver, heart, and lungs. The scan consisted of a 6-echo (TE$_1$=1.3 ms, $\Delta$TE=1 ms) gradient-recalled echo (GRE) protocol designed to minimize T1 bias (3° flip angle), Pixel Bandwidth = 1565 Hz, and 232 x 256 reconstructed image size, with 5 mm slice thickness and 1.72 x 1.72 mm$^2$ in-plane resolution.

## Overview of the Proposed Methodology

An overview of the proposed methodology is shown in Figure 3 and compared with current complex-based water-fat separation pipelines.

Current complex-based methods perform (1) regularized B$_0$ field mapping; (2) complex-based estimation using the estimated field map, that yields unique PDFF and $R_2^*$ maps; and (3) magnitude-based refinement of PDFF to correct phase errors. In 'mixed' magnitude/complex fitting[31], steps 2 and 3 are performed simultaneously, but are in practice initialized with a regularized field map from step 1.

For this work, single-step (i.e., non-iterative) graph-cut methods were used as the reference complex-based methods for validation and benchmarking. For the single-slice datasets, we used the 2-D method GlObally Optimal Surface Estimation (GOOSE)[15] for step 1 above, and Hybrid T2*-IDEAL[18] (hIDEAL) for steps 2 and 3. This method is referred to as GOOSE:hIDEAL. For the multi-slice dataset, we used vlGC:hIDEAL, by combining the 3-D method variable-layer graph-cut (vlGC)[14] for step 1, and hIDEAL for steps 2 and 3. Voxel-independent hIDEAL and iGC:hIDEAL –iterative graph-cuts (iGC)[8] plus hIDEAL– were also run for benchmarking.

In our methodology, PDFF and $R_2^*$ estimation are performed first by MAGO[20] using magnitude data only. Then, if phase data is deemed to be reliable, the proposed field mapping method may be used. To this point, field mapping may run independently at each voxel. This is referred to as *proposed voxel-independent method* throughout this work. The field map may be further adjusted and used to refine the magnitude-based PDFF and $R_2^*$ maps, yielding the *proposed adjusted method*. One simple adjustment step is described in the next section.

## Proposed Adjustment of Voxel-Independent Field Map and PDFF and $R_2^*$

Adjusting the voxel-independent field map enables the refinement of the PDFF and $R_2^*$ maps from MAGO. The voxel-independent field maps need to have been computed (using the *proposed voxel-independent field mapping* method) for both the chosen MAGO solution set and the alternative set, obtaining $f_B^c$ and $f_B^a$, respectively. Then, the adjusted field map may be created by selecting at each voxel $f_B^c$ or $f_B^a$, whichever is closest to a 'smoothed' version of $f_B^c$, $v(f_B^c)$. If $f_B^a$ and $v(f_B^c)$ are closer than $f_B^c$ and $v(f_B^c)$ are, then the chosen and alternative solution sets are swapped for that voxel. This procedure is similar to steps 2 and 3 of the RIPE (Regional Iterative Phasor Extraction) method, using a single iteration[32].

Note that this step yields an adjusted field map as well as adjusted PDFF and $R_2^*$ maps, because the whole solution set is consistently kept when swapping. In this work, median filtering with kernel size of 15-by-15 neighborhood was used to create $v(f_B^c)$.

The robustness of this adjustment step to noise was tested as described in the next section.

## Simulations

A slowly-varying synthetic field map was applied to one UK Biobank subject to explore the observation from Theory illustrated in Figure 2. The water magnitude, fat magnitude, and $R_2^*$ outputs from GOOSE:hIDEAL were used to synthesize complex-valued echo images with the synthetic field map and an arbitrary initial phase offset of $\pi/4$ rad for the whole image. The synthetic field map was set to vary linearly starting from the center of the image. The echo times from the original dataset were used in the simulation, and the 6-peak fat model. The synthesized complex-valued echo data was reconstructed with voxel-independent hIDEAL as well as MAGO followed by the proposed voxel-independent method.

Simulations were also performed to test the noise performance of the proposed field map adjustment step at refining the MAGO estimates. All UK Biobank subjects were used for this experiment. For each subject, additive complex Gaussian noise was progressively applied to the echo images. At each noise level, the Signal-to-Noise Ratio (SNR) was measured in the liver from the first echo magnitude image, by taking the ratio of the median signal in the liver and the median signal in the background. SNR was brought down to 5 for each subject. PDFF and field map were estimated at each noise level using (1) voxel-independent hIDEAL, (2) the proposed voxel-independent field mapping, and (3) the proposed adjusted method with PDFF refinement. For each noise level (and each method), the percentage of 'swapped' liver voxels was computed by comparing the output PDFF map with a reference PDFF reconstructed with GOOSE:hIDEAL using the original data with no added noise. 'Swapped' voxels were defined according to Berglund and Skorpil[13], i.e. those with an absolute PDFF difference higher than 10% and with the opposite dominant species compared to the reference PDFF.

## In-Vivo Evaluation

The proposed voxel-independent method and the proposed adjusted method were demonstrated in the 1.5 T UK Biobank datasets and in the 3 T multi-slice dataset.

The 152 UK Biobank subjects were used to validate the proposed methods for *in vivo* agreement with GOOSE:hIDEAL under low $B_0$ inhomogeneities. Voxel-independent hIDEAL and iGC:hIDEAL were also run for comparison. Median field map, PDFF and $R_2^*$ values within the liver segmentation masks were extracted for all methods and compared using Bland-Altman analysis[33]. For this comparison, in order to remove differences arising from different optimization settings, the output field maps of the proposed methods were used to initialize hIDEAL to compute PDFF and $R_2^*$.

The multi-slice 3 T dataset was reconstructed with the proposed methods and GOOSE:hIDEAL (both 2-D based) and with the 3-D method vlGC:hIDEAL for reference. Bipolar gradients' correction was run for the proposed methods as described in Theory.

# Results

## Simulations

Figure 4 shows the subject with the additional synthetic field map, reconstructed using voxel-independent hIDEAL and the proposed voxel-independent method. In accordance with the observations from Figure 2, voxel-independent hIDEAL erroneously converges to an aliased solution when the initial field map value (0 Hz) is not within the true field $\pm\Delta f/2$ (with $\Delta f \approx 220$ Hz at 1.5T). This results in fat-water swaps. Consistently with Figure 2 also, the proposed voxel-independent method enables accurate field map estimation beyond the $\pm\Delta f/2$ range, up to slightly beyond $\pm\Delta f$ (approximately $\pm 250$ Hz). Note the field map wraps in the proposed method do not affect PDFF, since field map estimation is performed after PDFF mapping, unlike in the complex-based case, where PDFF is determined by the field map.

Figure 5 shows the noise performance of the proposed field map adjustment step on the UK Biobank population, where each subject was reconstructed under progressive noise, expressed as the percentage of swapped voxels. The adjusted method keeps the percentage low for the SNR levels tested. When taking summary metrics such as the median, a small number of swapped voxels have little contribution to the reported value (see In-Vivo Evaluation for quantification results). Conversely, the proposed voxel-independent implementation exceeds 5% median at SNR near 14. Overall, for SNR in typical abdominal six-echo relaxometry, the adjustment step shows modest improvement over voxel-independent results.

Figure 6 and Figure 7 show the field map adjustment procedure with refinement of the MAGO PDFF map for a simulated SNR of 10 in two UK Biobank subjects, one with liver PDFF median = 5.1% and one with liver PDFF median = 18.0%. The field map adjustment step is able to 'unswap' the majority of voxels, rendering a solution that is comparable to the reference PDFF computed using GOOSE:hIDEAL on the original data with no added noise.

## In-Vivo Evaluation

Table 1**Error! Reference source not found.** shows results from Bland-Altman analysis of median field map, PDFF and $R_2^*$ within the liver for various methods in the UK Biobank subjects. All methods (voxel-independent hIDEAL, iGC, proposed voxel-independent, proposed adjusted) were compared to GOOSE:hIDEAL. For field map quantification, very small bias and very low variability were observed for the proposed adjusted method, with overall excellent agreement with the reference. For PDFF and $R_2^*$ quantification, excellent agreement was observed between the proposed adjusted method and GOOSE:hIDEAL. Voxel-independent hIDEAL showed higher bias of 0.6 Hz and higher variability of [-5.8, 7.0] Hz and comparable bias but higher limits of agreement of [-0.2, 0.3] % for PDFF and [-0.3, 0.3] Hz for $R_2^*$.

Figure 8 shows one subject from the UK Biobank reconstructed using four methods: voxel-independent hIDEAL, GOOSE:hIDEAL, proposed voxel-independent, and proposed adjusted. Note that field mapping follows PDFF estimation for the proposed methods. The proposed methods produce comparable results to GOOSE:hIDEAL for the observed inhomogeneity range. Voxel-independent hIDEAL suffers from field map wraps that cause fat-water swaps on PDFF including 'double swaps' within the liver and spleen (double swaps yield plausible PDFF values, which may lead to misdiagnosis if missed[7]), and inaccurate $R_2^*$. The proposed adjusted method is able to unswap a few voxels within the liver that were misidentified using the proposed voxel-independent method.

Figure 9 shows seven slices from the 3 T dataset with liver coverage, processed using the proposed adjusted method and GOOSE:hIDEAL (both 2-D-based) and using the reference vlGC:hIDEAL (3-D-based). Magnitude images from the first echo time are included for anatomical reference. The scan shows both overall high field inhomogeneities as well as rapid field transitions near edges of organs, for example the dome of the liver and the spleen. Though no ground truth maps are available, the slice-by-slice variation in field map ought to be consistent, with increasing inhomogeneity further away from the isocenter. Also, wrong field map convergence can be identified when fat-water swaps are observed in the corresponding locations of the water-separated maps. The field maps by the proposed method are consistent across slices and generally in agreement with the 3-D method. GOOSE:hIDEAL presented fat-water swaps within the subject's left arm that were not observed in vlGC:hIDEAL or in the proposed method. No substantial fat-water swaps were observed for the proposed method.

Figure S1 and Figure S2 in Supplementary Material show reconstructions of the multi-slice 3 T dataset using other widely used methods from the literature, namely Region Growing[7], iGC[8] and B0-NICE[22]. All of these implementations returned at least one slice with a whole-liver fat-water swap (Fig. S2) resulting from wrong field map convergence (Fig. S1). Results from the proposed voxel-independent method are also included. These results help in judging the challenges of field map reconstruction on the 3 T dataset from this work, and also benchmark the proposed method among those in the literature.

## Discussion

This work demonstrated the feasibility of field mapping following ambiguity-resolved magnitude-based water-fat separation (MAGO). The proposed method is efficient and relatively simple to implement. The method in its voxel-independent form doubled the traditional range limit of voxel-independent complex-based methods. Simulations demonstrated robustness to noise of the proposed adjustment of the method. Field map, PDFF and $R_2^*$ quantification were validated against a complex-based reference in simulated data and 152 subjects from the UK Biobank. The method accurately reconstructed a 3 T dataset with field inhomogeneities, including rapid transitions near organ structures, showing superior performance to a 2-D-based reference and comparable performance to a 3-D-based reference.

The PDFF map and the $R_2^*$ map produced by our method are initially magnitude-based. The complex-valued water and multi-spectral fat signal terms are then derived and demodulated from the complex-valued raw data. This removes local minima associated with the phase shift due to the off-resonance frequency of fat when using the proposed field mapping. For clinical scanners where obtaining reliable phase images is challenging, our methodology still enables consistent PDFF and $R_2^*$ estimation because it is run first via MAGO[20]. For this reason, our pipeline may be standardized and deployed across a range of scanning conditions, including routinely observed SNR at both clinical field strengths (1.5 T and 3 T), as shown by the data used in this work.

Smoothness constraints commonly used by field map regularization in complex-based methodology may not hold true in regions of rapid field variation, and may propagate wrong field map and PDFF estimates to the organ of interest. Regularization methods that are initialized using seed pixels in down-sampled data will be particularly sensitive[7,9]. Slices distanced from the isocenter or closer to the lungs tend to present higher field inhomogeneities, as illustrated by the 3 T dataset. The proposed method produced accurate field maps under the observed field inhomogeneity range in-vivo, which exceeded 400 Hz at 3 T; however, further validation at high fields is needed.

While the proposed field map adjustment may be considered a form of spatial regularization, note that it is subtly distinct, because the postprocessed field map at a given voxel is chosen from one of the two pre-computed candidate MAGO solutions. Thus, the field map value at a given voxel is derived directly from that voxel's complex-valued raw data, but it is 'chosen' considering neighboring information. This avoids issues from some previously used spatial regularization, where the field map is forced to assume a value within a predefined range, for example by neighborhood interpolation in multiresolution techniques[7]. Our approach did not penalize rapid field variations and preserved a high-frequency appearance of the field map, where other methods may yield over-smoothed results.

Note that the field map adjustment step may lead to a chosen MAGO solution set that is not necessarily associated with the lower fitting residual. Instead, the solution most consistent with the voxel neighborhood is chosen. This is useful where noise corrupts the raw data magnitude, causing the physiologically correct solution to become a local optimum. For comparison purposes, for PDFF and $R_2^*$ quantification the adjusted field maps were used to initialize hIDEAL; in practice, this is undesirable as it may propagate phase errors, which in turn needs magnitude-based refinement. This is the argument for hybrid fitting[18]. Choosing from the two candidate magnitude-based solutions removes the need for an additional magnitude-based refinement step.

This work had several limitations. Our method's reported range limit could limit its use in applications like QSM. However, regularization techniques may be used to extend this range, and our method can provide useful initialization and additional information to regularizers. Additionally, multi-slice data can benefit from additional constraints applied in the slice direction, as demonstrated previously[13,14,16,34]. Disadvantages of 3-D methods include larger processing times (e.g. in order to build the graph) and memory requirements, which impairs real-time reconstruction; if the volume is subdivided into multi-slice subsets, results for a given slice may change depending on how many slices are used by the method at once. Note that any multi-slice extensions of the method could be applied to the 3 T dataset from this work, though median filtering may need to be revisited in favor of piece-wise techniques.

While two different six-echo acquisition protocols were evaluated, the $\Delta TE$ was approximately half at 3 T, where the chemical shift is double; therefore, the two protocols can be expected to result in similar method performance. The sensitivity of the proposed method to acquisition parameters needs to be evaluated, though MAGO has been previously evaluated using different acquisition protocols[20].

# Conclusions

Our results suggest that magnitude-based ambiguity-resolved water, fat and $R_2^*$ maps may be used to estimate a field map using the complex-valued raw data, so that $B_0$ field mapping need not be a critical step for correct water-fat separation, but rather improves proton density fat fraction estimates. This approach for full-range PDFF and field mapping is fast and widely applicable, and may be useful initialization for previously proposed field map regularization techniques.

# List of Figures

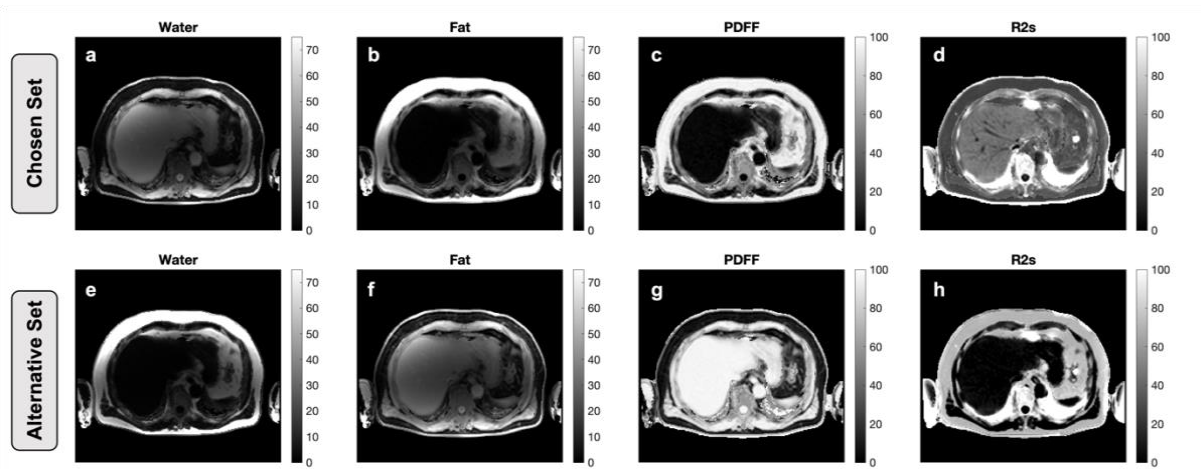

Figure 1. Illustration of magnitude-only fitting with resolved water-fat ambiguity (MAGO) on a UK Biobank subject. MAGO uses multipoint search to converge to two solution sets, the water-dominant set and the fat-dominant set, where each set contains amounts of water, fat, PDFF and R2*. Then chooses, at each voxel, the solution with lowest fitting residual, yielding a chosen set (a-d) and an alternative set (e-h). This enables PDFF estimation over the full range (0 to 100%) as well as correct $R_2^*$ estimation. MAGO, magnitude-only fitting; PDFF, proton density fat fraction

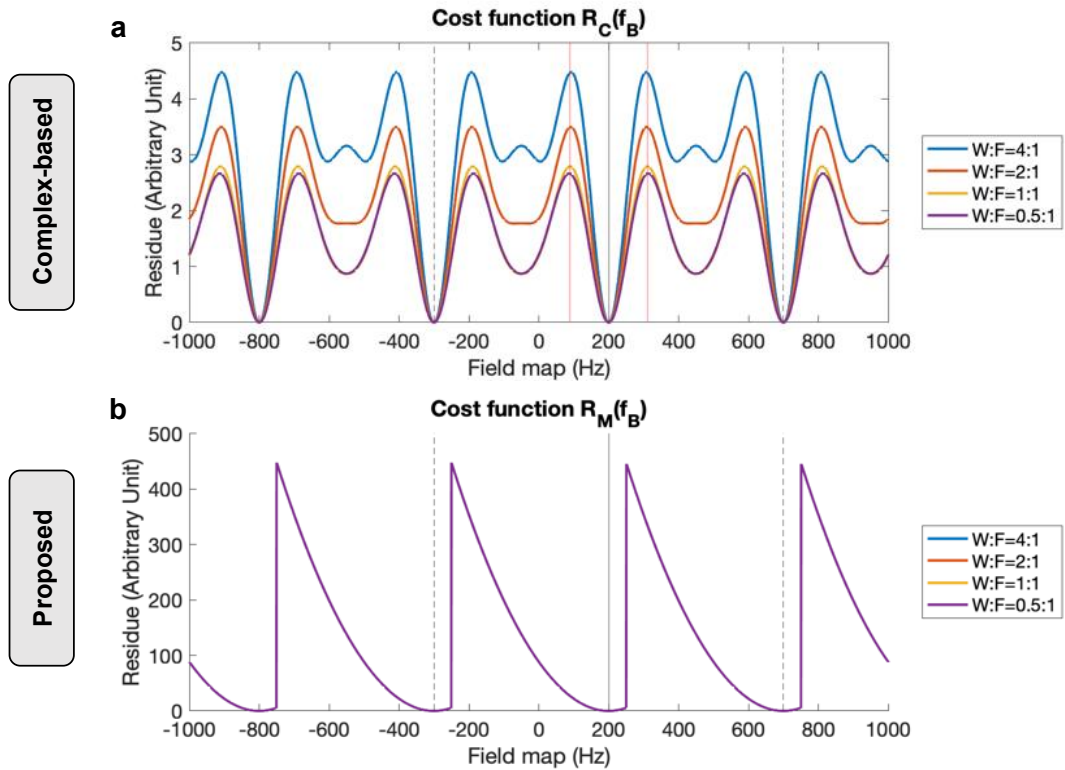

Figure 2. Simulated loss functions for the complex-based estimation problem (a) and for the proposed formulation (b) within a voxel. True field map = 200 Hz (solid grey line), varying voxel Water:Fat proportion, $R_2^* = 30$ Hz, $\phi_0 = 0$ rad, echo times = [2, 4, 6] ms, 1.5 Tesla, 6-peak liver fat spectrum. Voxel-independent complex-based estimation (e.g. IDEAL) converges to an incorrect local minimum near 0 Hz, whereas the proposed voxel-independent method converges to the true value. The proposed loss function is also independent of Water:Fat proportion within the voxel. The period, ±1/ΔTE, is marked (dashed lines). True solution ± $\Delta f/2$ is marked (red lines in a). IDEAL, iterative decomposition of water and fat using echo asymmetry and least squares estimation

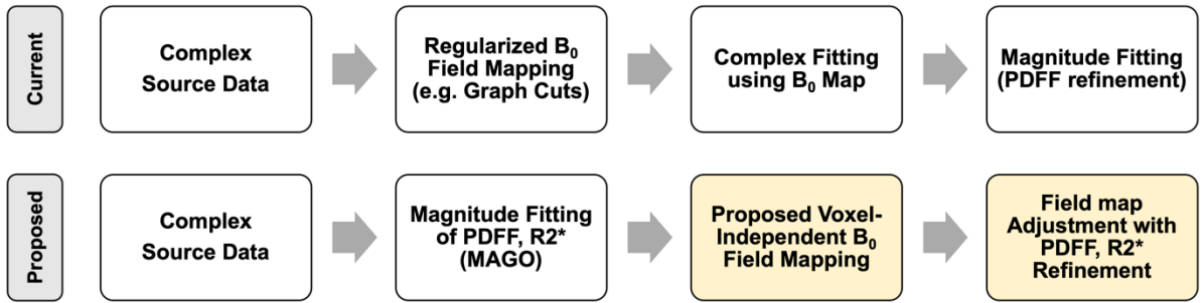

Figure 3. Current (top) and proposed (bottom) methods for field mapping and PDFF estimation, given complex-valued source data. The yellow modules are those demonstrated in this work. Current methods perform field mapping with regularization first, then derive PDFF and $R_2^*$, and include a final magnitude-based refinement step. The proposed method applies magnitude fitting with resolved water-fat ambiguity (MAGO) first, then performs voxel-independent field mapping, and includes a final adjustment step to refines the MAGO PDFF and $R_2^*$.

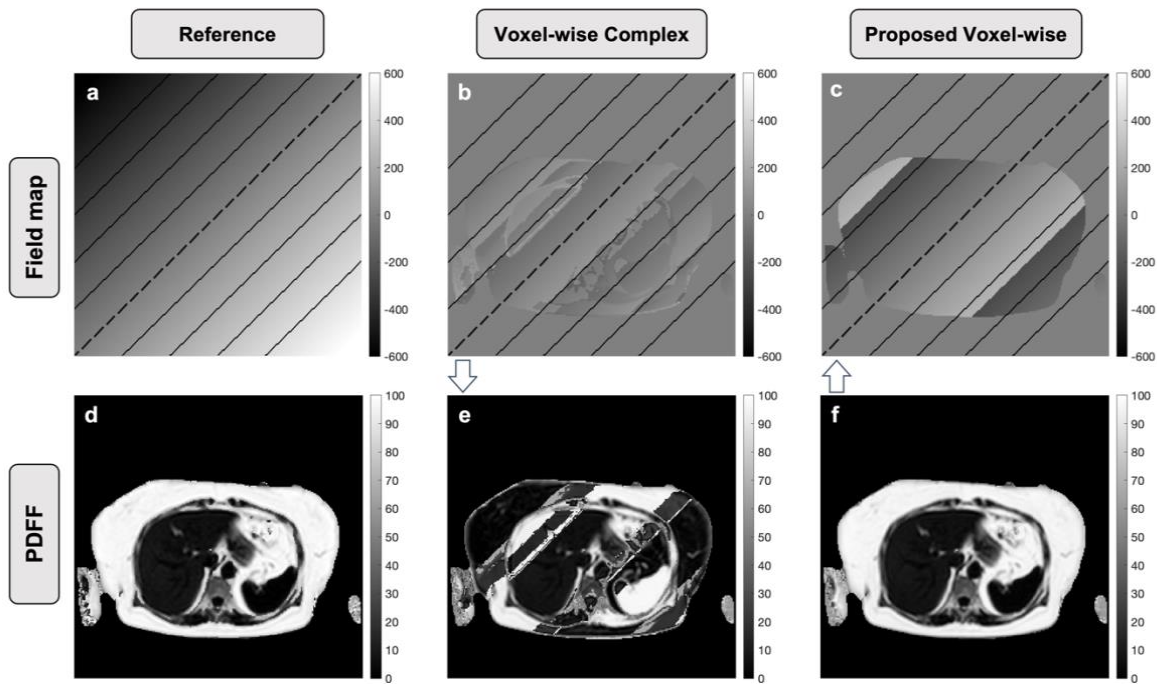

Figure 4. UK Biobank subject with a synthetic field map added. The reference PDFF (a) was obtained using regularized complex-based estimation by GOOSE:hIDEAL prior to adding the synthetic field map (a). Reference field map equal to zero (dashed line, a-c) and reference field map equal to $\pm\Delta f/2$ (solid lines, a-c) are indicated. Voxel-independent hIDEAL (b, e) misconverges beyond $\pm\Delta f/2$. The proposed voxel-independent implementation (c, f) is accurate up to beyond $\pm\Delta f$. Note the wraps in the proposed field map do not affect PDFF, since field map estimation is performed after PDFF is obtained, unlike complex-based methods (dependence indicated by white arrows). GOOSE, fat water decomposition using globally optimal surface estimation; hIDEAL, hybrid IDEAL

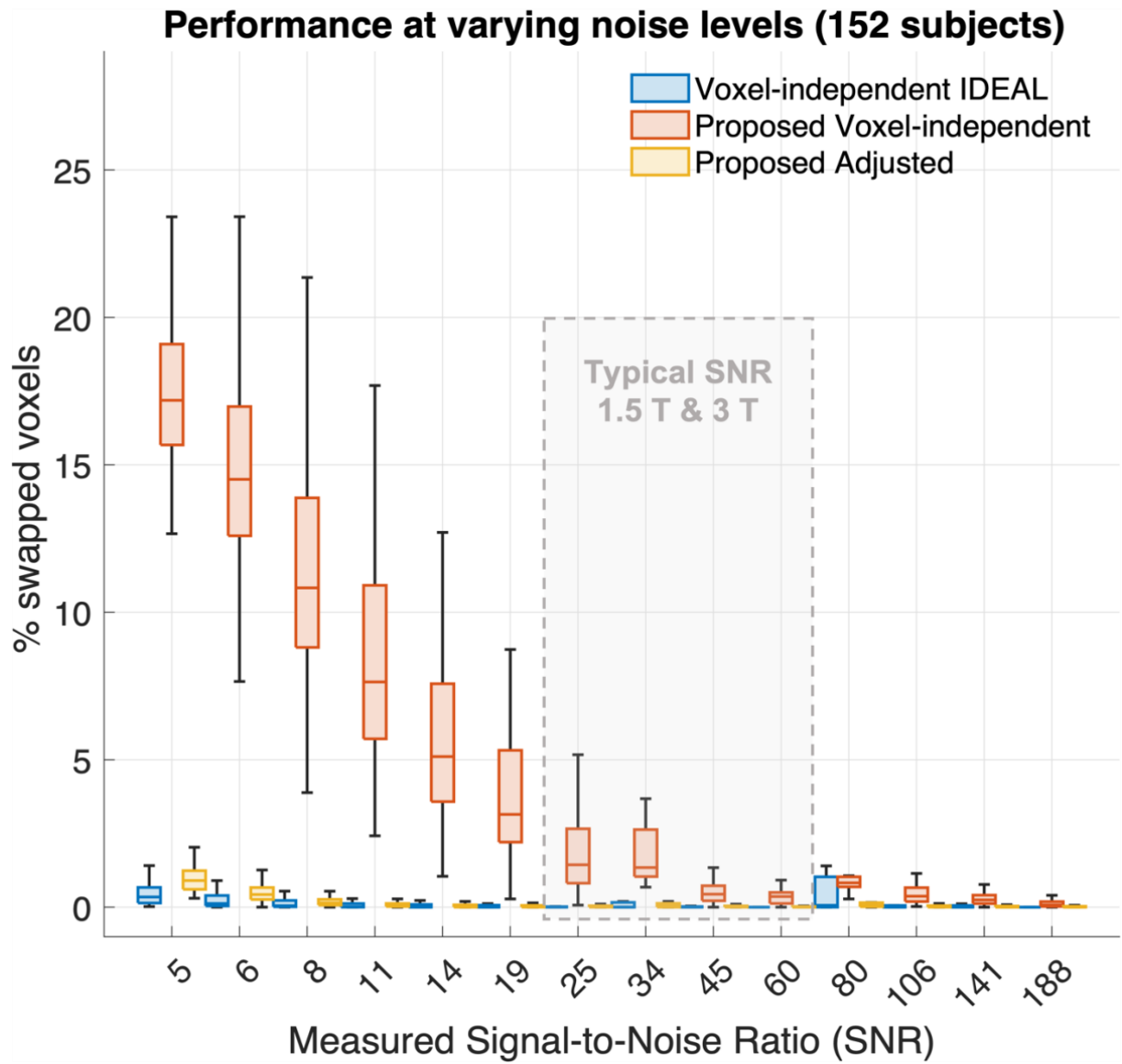

Figure 5. Fraction of swapped voxels as a function of SNR in the UK Biobank population (N=152). For all subjects, noise was progressively added to the raw data and reconstructed using voxel-independent IDEAL (blue), the proposed voxel-independent method (red) and the method with the proposed adjustment. The number of swapped voxels was determined by comparing the reconstructed PDFF for each method against a reference map computed on the original data with no added noise. SNR was measured within the liver for each noise step. Typical SNR in six-echo relaxometry at 1.5 T and 3 T are represented by the shaded box. Outliers were omitted during plotting only. SNR, signal-to-noise ratio

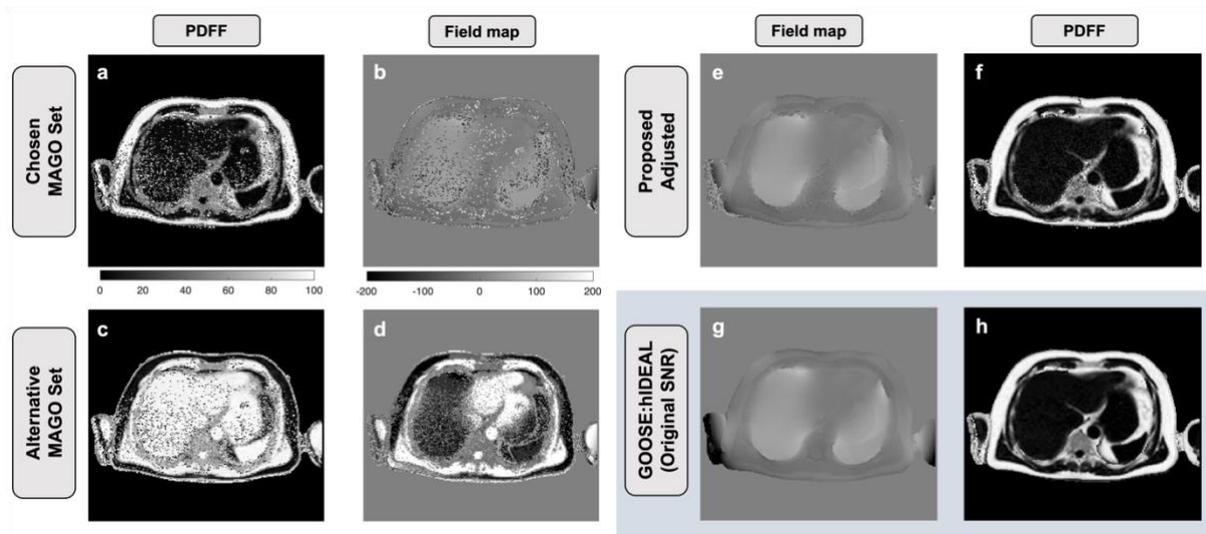

Figure 6. Proposed field map adjustment step on a UK Biobank subject after noise was added to the raw data, yielding SNR=10 within the liver. Results for the proposed voxel-independent method are shown (a, b), corresponding to the chosen MAGO solution set. The proposed adjustment step chooses at each voxel from the chosen MAGO solution (a, b) and the alternative MAGO solution (c, d) by comparing them to a smoothed version of the chosen field map (b). The resulting adjusted field map (e) and PDFF (f) have most voxels corrected using this approach. Reference maps computed by GOOSE:hIDEAL using the original data with no added noise are included for comparison (g, h).

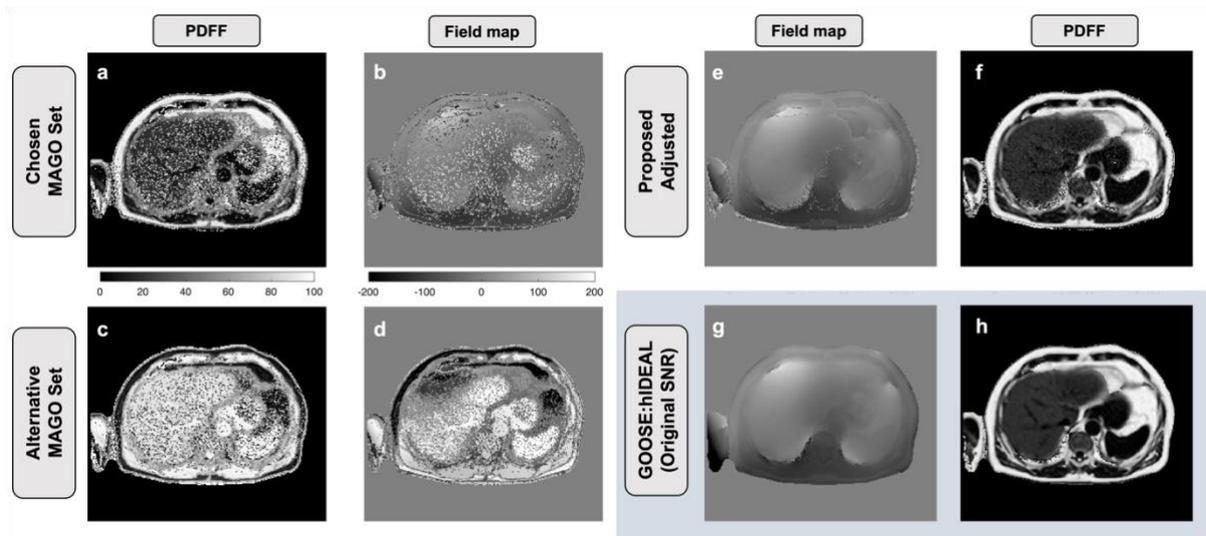

Figure 7. Proposed field map adjustment step on a UK Biobank subject after noise was added to the raw data, yielding SNR=10 within the liver. Results for the proposed voxel-independent method are shown (a, b), corresponding to the chosen MAGO solution set. The proposed adjustment step chooses at each voxel from the chosen MAGO solution (a, b) and the alternative MAGO solution (c, d) by comparing them to a smoothed version of the chosen field map (b). The resulting adjusted field map (e) and PDFF (f) have most voxels corrected using this approach. Reference maps computed by GOOSE:hIDEAL using the original data with no added noise are included for comparison (g, h).

| N=152 | Field map (Hz) | PDFF (%) | R2* (Hz) |
|---|---|---|---|
| **Voxel-Independent IDEAL** | 0.594 [-5.79, 6.97] | 0.003 [-0.24, 0.25] | -0.021 [-0.29, 0.25] |
| **iGC** | 0.004 [-0.10, 0.10] | **-0.000 [-0.01, 0.01]** | 0.002 [-0.04, 0.04] |
| **Proposed Voxel-independent** | 0.014 [-0.34, 0.37] | -0.000 [-0.02, 0.02] | 0.016 [-0.05, 0.08] |
| **Proposed Adjusted** | **-0.000 [-0.01, 0.01]** | **0.000 [-0.01, 0.01]** | **-0.000 [-0.00, 0.00]** |

Table 1. Methods were compared against GOOSE:hIDEAL on the UK Biobank population (N=152) using Bland-Altman analysis. Each method's field map was used to initialize hIDEAL for PDFF and R2* estimation. Values are expressed as 'bias [lower LoA, upper LoA]'. vlGC, variable-layer graph-cut based method; LoA, limits of agreement

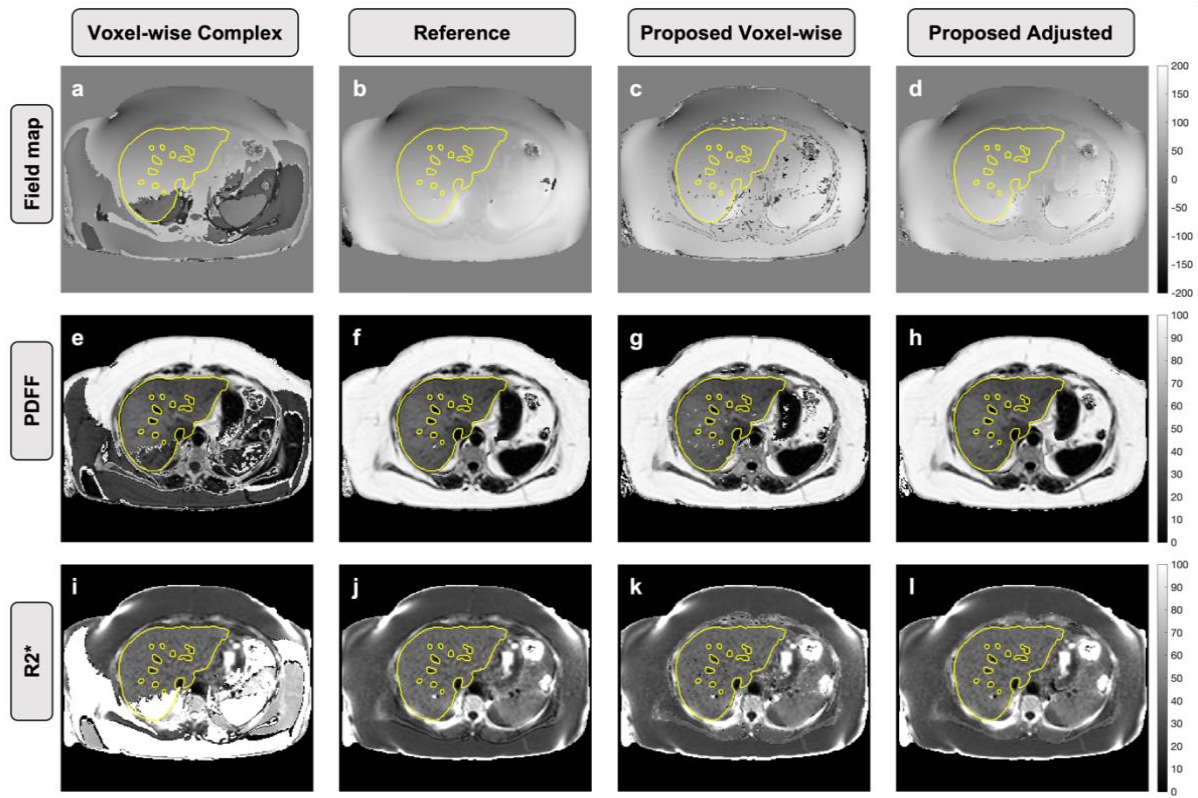

Figure 8. The subject from UK Biobank with highest PDFF disagreement between reference GOOSE:hIDEAL and voxel-independent hIDEAL is shown, reconstructed using GOOSE:hIDEAL (a, e, i), voxel-independent hIDEAL (b, f, j), proposed voxel-wise method (c, g, k) and proposed adjusted method (d, h, l). Liver segmentations are outlined (bright contours). Voxel-independent hIDEAL presents field map wraps which lead to fat-water swaps within the liver and subcutaneous fat, as well as inaccurate R2*. The proposed adjusted method is able to unswap a few voxels within the liver that were misidentified using the proposed voxel-independent method.

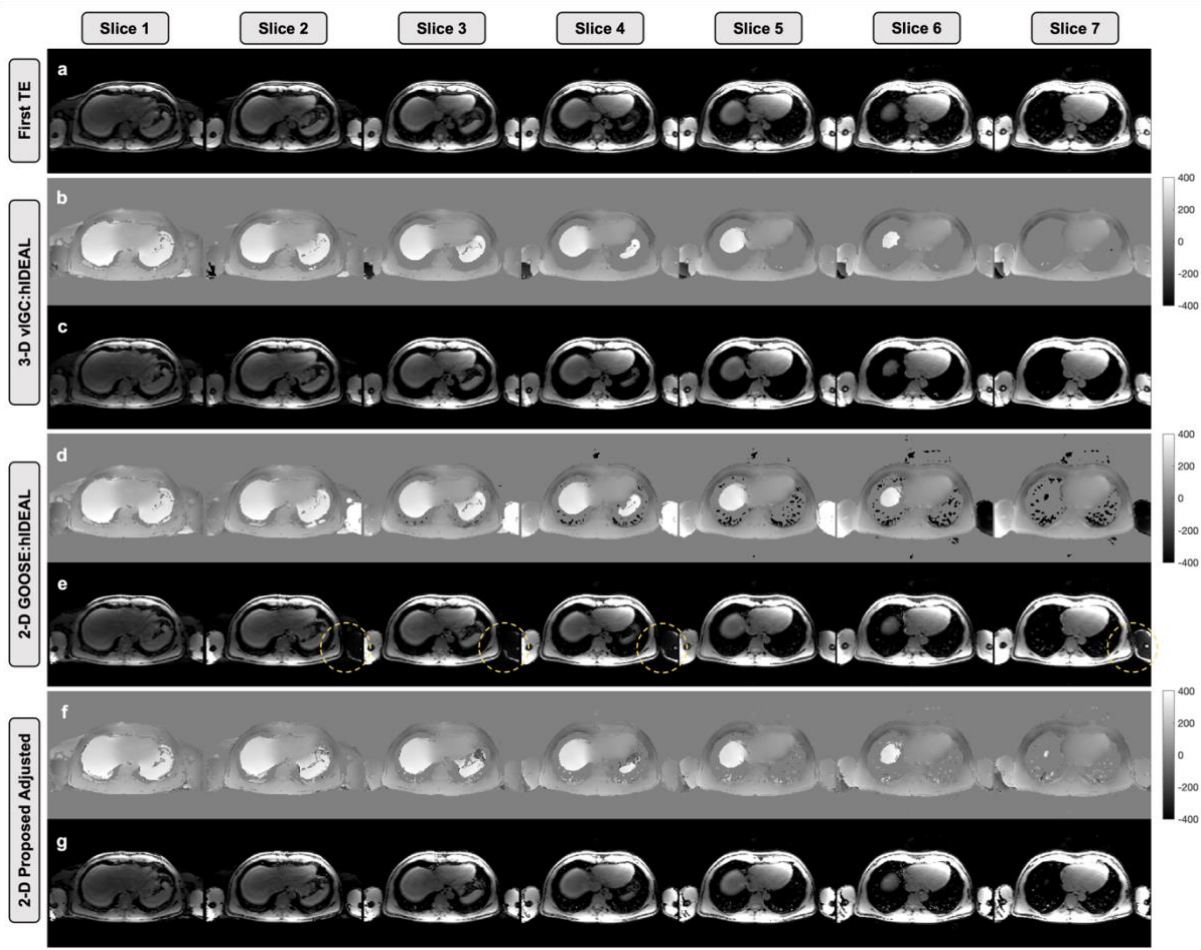

Figure 9. Output field map and water-separated images from seven slices of the 3 T dataset with liver coverage. Magnitude images from the first echo time are provided for anatomical reference (a). The slices were processed using 3-D vlGC:hIDEAL (b, c), 2-D GOOSE:hIDEAL (d, e), and the 2-D proposed adjusted method (f, g). High field inhomogeneities and rapid field transitions are observed, especially near the dome of the liver and the spleen. The methods were generally in agreement, but GOOSE:hIDEAL presented fat-water swaps within the subject's left arm (bright dashed circles in e) that were not observed in vlGC:hIDEAL nor in the proposed adjusted method.

# Supplementary Material

The 3 T dataset was reconstructed using other widely used methods from the literature for comparison: Region Growing[1], iterative Graph Cuts (iGC)[2], B0-NICE[3], as well as the proposed voxel-independent method and the proposed adjusted method.

The authors acknowledge the use of the Fat-Water Toolbox for the Region Growing and iGC results shown. The Toolbox is available under http://ismrm.org/workshops/FatWater12/data.htm. B0-NICE is available under https://www.mathworks.com/matlabcentral/fileexchange/48313-b0-mapping-b0-nice.

Region Growing from the Fat-Water Toolbox was run with default parameters, but the raw data x-dimension was padded to enable down-sampling by an exact factor of 4. Iterative Graph Cuts from the Fat-Water Toolbox was run with default parameters except for field map search range that was increased from [-400, 400] Hz to [-600, 600] Hz. B0-NICE was run with default parameters.

Figure S1 and Figure S2 show the field maps and water-only images output from the different methods.

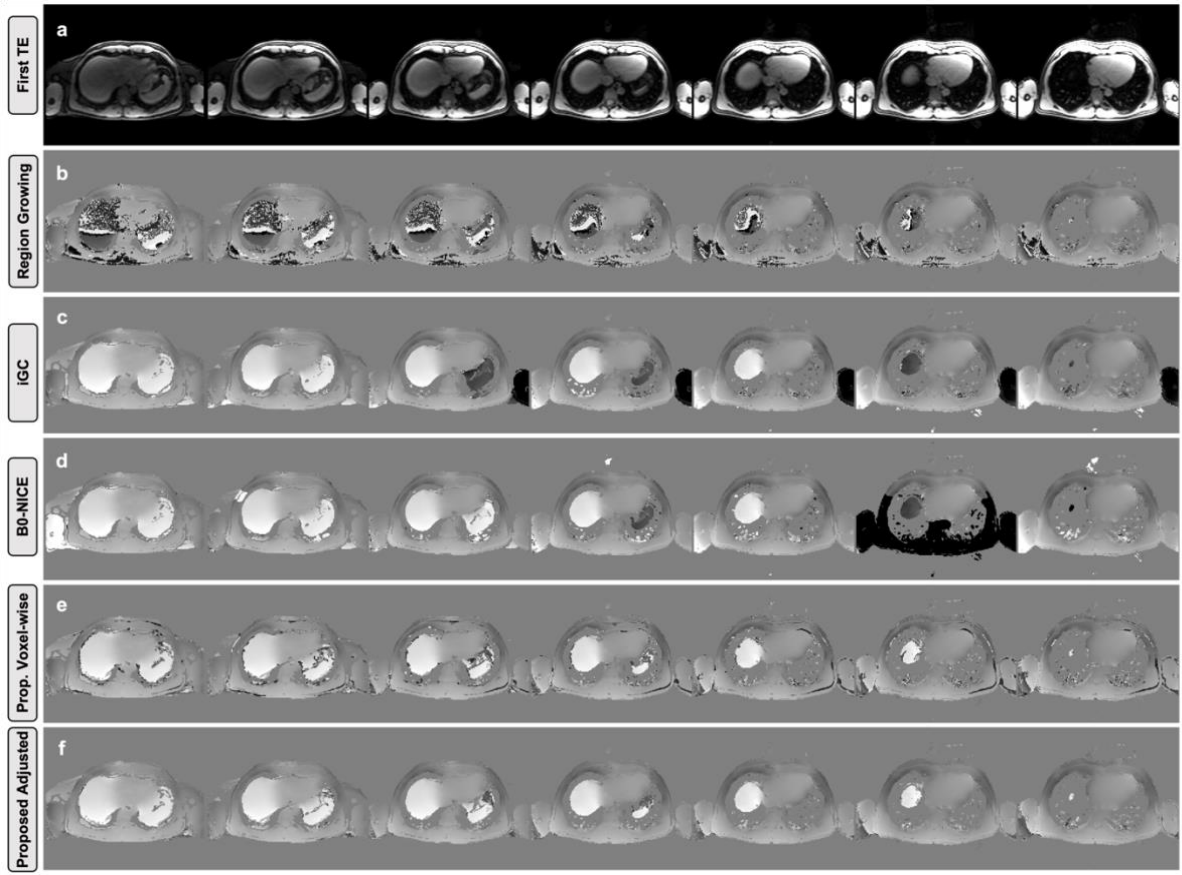

Figure S1. Field maps reconstructed using several widely used methods from the literature. Magnitude images of the first echo time are included for each slice for anatomical reference (a). Methods used were Region Growing (b), iterative Graph Cuts (c), B0-NICE (d), Proposed voxel-independent method (e), Proposed adjusted method (g).

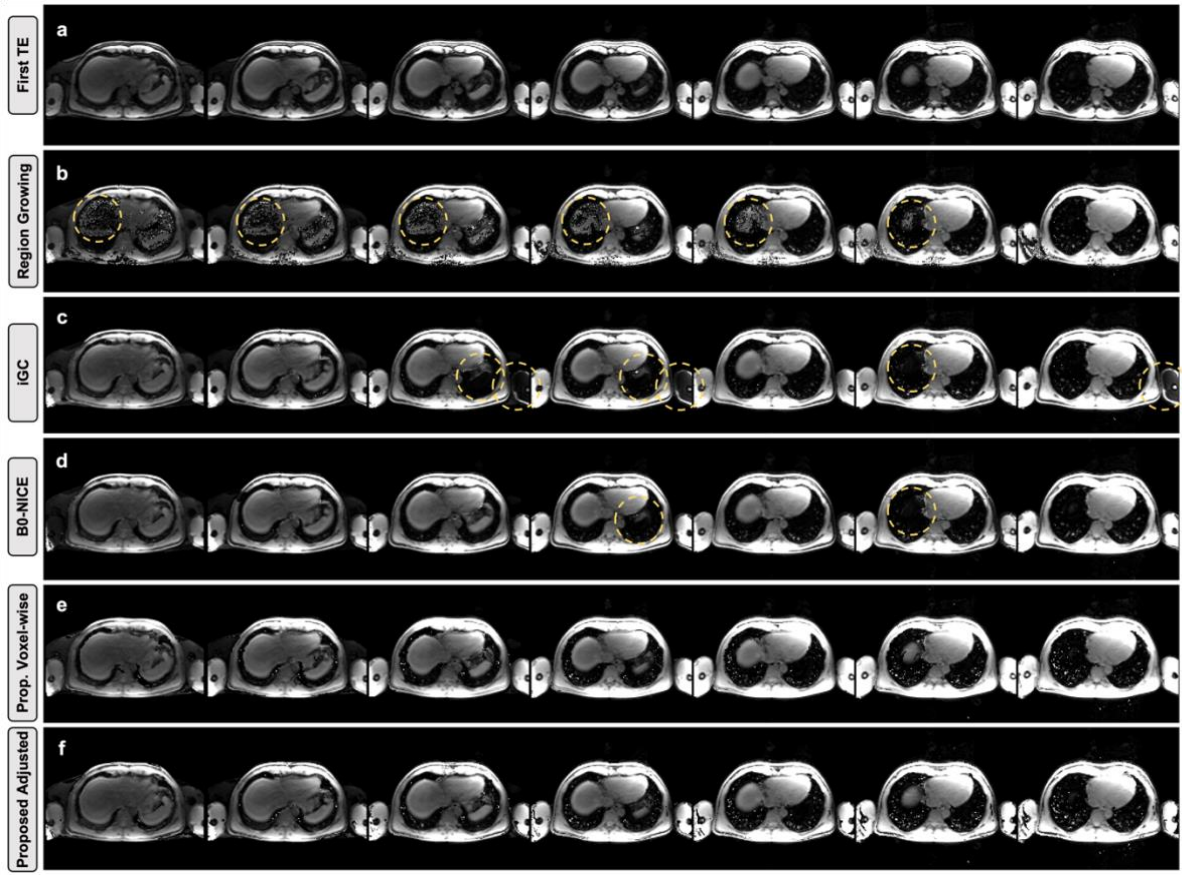

Figure S2. Water-separated maps reconstructed using other widely used methods from the literature. Magnitude images of the first echo time are included for each slice for anatomical reference (a). Methods used were Region Growing (b), iterative Graph Cuts (c), B0-NICE (d), Proposed voxel-independent method (e), Proposed adjusted method (g). Major fat-water swaps are indicated (bright dashed circles).